# Metaphysics of the Principle of Least Action[1]

Vladislav Terekhovich[2]

**Abstract**

Despite the importance of the variational principles of physics, there have been relatively few attempts to consider them for a realistic framework. In addition to the old teleological question, this paper continues the recent discussion regarding the modal involvement of the principle of least action and its relations with the Humean view of the laws of nature. The reality of possible paths in the principle of least action is examined from the perspectives of the contemporary metaphysics of modality and Leibniz's concept of essences or possibles striving for existence. I elaborate a modal interpretation of the principle of least action that replaces a classical representation of a system's motion along a single history in the actual modality by simultaneous motions along an infinite set of all possible histories in the possible modality. This model is based on an intuition that deep ontological connections exist between the possible paths in the principle of least action and possible quantum histories in the Feynman path integral. I interpret the action as a physical measure of the essence of every possible history. Therefore only one actual history has the highest degree of the essence and minimal action. To address the issue of necessity, I assume that the principle of least action has a general physical necessity and lies between the laws of motion with a limited physical necessity and certain laws with a metaphysical necessity.

## 1 Introduction

The principle of least action (PLA) is one of the most general laws of theoretical physics and simultaneously one of the most philosophically conflicting laws. Over the centuries, many scientists have linked it to hopes of a universal theory, despite the related metaphysical disputes about causality. Fermat, Leibniz, Maupertuis, and Euler were sure that nature is thrifty in all its actions thanks to the perfection of God. Planck believed that, "among the more or less general laws which manifest the achievements of physical science in the course of recent centuries, the Principle of Least Action is probably the one, which, as regards form and content, may claim to come nearest to that final ideal goal of theoretical research" (Stöltzner, 2003).

The PLA and other variational or extremal principles provide an alternative and more global approach to mechanics than Newton's laws. The PLA and the calculus of

---

variation, in general, are more global than local differential equations and are widely used for solving dynamic tasks in diverse fields of physics such as classical mechanics, electrodynamics, relativity theory, and quantum physics. In the PLA, the action is the integral of a certain expression along a possible path or history of a system in a configuration space. The expression can be Lagrangian, for instance, the difference between kinetic and potential energy or, in the case of continuous fields, the Lagrangian density. The integral can be over the path, time, n-dimensional volume, or four-dimensional space-time. In the quantum field theory, the action has a meaning of the phase of quantum amplitude (Feynman & Hibbs, 1965). In other variational principles, many characteristics take a minimal or maximal value from all possible values. These could include: the optical length, constraint, proper time, curvature of space-time, and thermodynamic potentials (Lanczos, 1986; Landau & Lifshitz, 1975; Stöltzner, 1994; Lemons, 1997; Goldstein et al., 2002; Yourgrau & Mandelstam, 1968; Taylor & Wheeler, 2000; Papastavridis, 2002; Sieniutycz & Farkas, 2005; Hanc & Taylor, 2004; Ogborn & Taylor, 2005).

And yet, the PLA has always been surrounded by a fog of mysticism. The system seems to "choose" the actual path along which an action is less than along other paths. It is as if the system's final state determines the path that the system takes to reach that state. On the one hand, we cannot allege that an object actually "chooses" or "calculates" the path of minimal action. On the other hand, it appears that the actual path is somehow connected with the future actual state or event. A general principle of causality states that a cause should always precede its effect. This view of causality is used in most of the physical laws and is consistent with the grounded belief that causal influences cannot travel backwards in time. Nevertheless, until today, we have not understood how a physical system seems to "choose" an actual path or history from all possibilities for motion or why this actual history involves minimal action. Moreover, the history of physical teleology might alternatively suggest a relationship between the PLA and the problem of determinism (Stöltzner, 2003). Besides being between teleology and determinism, the PLA takes a special place among other physical laws.[3] Additionally, it appeals to a modal notion of "possibilities".

Today, in spite of Planck's hope, the PLA is generally accepted only as a mathematical tool equivalent to the differential equations of motion (Yourgrau &

---

[3] As shown, the conservation laws can be derived from the variational principles (Goldstein, et al., 2002; Hanc & Taylor, 2004; Brizard, 2008).

Mandelstam, 1968, p. 178f). However, some physicists have tried to clarify the foundations of the variational principles (Polac, 1959; Asseev, 1977; Lanczos, 1986; Stöltzner, 1994, 2003; Yourgrau & Mandelstam, 1968; Wang, 2008). At the same time, as Butterfield (2004a) stressed, this topic seems wholly ignored in the philosophical literature about variational principles. He assumed that, thanks only to the rise of modal metaphysics in analytical philosophy, the topic is plainly visible nowadays and focused almost entirely on the way of specifying final conditions and teleology. Indeed, recently, some authors have examined how the PLA and other variational principles are involved in modal metaphysics (Butterfield, 2004a, 2004b; Katzav, 2004, 2005; Ellis, 2005; Bird, 2007; Terekhovich, 2013; Thebault & Smart, 2013). They have considered the relations between the PLA and causality, dispositional essentialism, the Humean view of the laws of nature, and the truthmaker principle. However, the study of modality is comprehensive and concerns some other issues connected with necessity and possibility. The themes of the modality and the nature of possible worlds are widely discussed in modal metaphysics (Plantinga, 1974; Adams, 1974; Kripke, 1980; Lewis, 1986; Chihara, 1998; Armstrong, 2004; Fine, 2005) and in relation to different physical phenomena (Shoemaker, 1984, Ellis, 2001; Bird, 2006).

This paper continues the recent discussion of the metaphysical issues of the PLA, especially regarding the modal involvement of the PLA. I think that Butterfield (2004a, 2004b) is right in that the whole analytical mechanics is steeped in modality. I am, moreover, sure that the most promising direction for the PLA is a metaphysical investigation of the possible paths or histories connected with the laws of quantum systems. In addition to presenting criticism of other concepts, I propose a positive solution for the metaphysical content of the PLA. First of all, I examine the question of a reality of "possible paths" or "possible histories" in the PLA, as well as how they are connected with the notion of "possible objects" or "possibilia" of the contemporary metaphysics of modality and of Leibniz's concept of the essences or possibles striving for existence.

This paper's solution for some of the metaphysical issues of the PLA is based on the intuition that quantum mechanics might be a key to understanding the philosophical content of this principle. I assume that deep ontological connections exist between the possible paths of the PLA and quantum possible histories of the Feynman path integral formalism (FPI).[4] This paper introduces a model of a two-level modality based on a realistic approach to the possible or virtual motions in the calculus of variations. It considers the possible paths in the Feynman integral to being descriptions of similar processes taking place in the possible modality of being. I elaborate the modal interpretation of the PLA that replaces the classical representation of the system's motion along a single history in the actual modality by the simultaneous motions along an infinite set of all possible histories in the possible modality. To address the issue of necessity, I assume the PLA to have a general physical necessity and to lie between the laws of motion (with a limited physical necessity) and certain laws (with a metaphysical necessity) that govern the PLA.

The rest of this paper is structured as follows. Section 2 gives a short description of the PLA. Section 3 illustrates the connection between the PLA and the FPI of quantum mechanics. Section 4 briefly introduces some metaphysical difficulties of the PLA related to causality, necessity, and a notion of possibility. Section 5 discusses the problem of the reality of the possible histories, possible objects, and possible worlds from the perspectives of various stances of modal metaphysics, including the Leibniz concept. The basic notions of the modal interpretation of the PLA are formulated in Section 6. The relations between the PLA and dispositional essentialism are considered in Section 7. Section 8 explains how the modal interpretation of the PLA can change the view of causality in the PLA. Section 9 compares the arguments in the debate regarding the Humean and non-Humean views of the laws of nature with concern to the PLA. Section 10 presents the paper's conclusions.

---

[4] It is known that the PLA is connected with the FPI through the notion of "action" and can be derived from the FPI as a limit on a large scale. Feynman even argued that the relation between symmetry laws and conservation laws is connected with the principle of least action "because they come from quantum mechanics" (Feynman, 1985, p. 105). Some authors have argued that all of classical mechanics could be represented as a short-wave approximation of quantum mechanics, and therefore, the action has the meaning of the phase of quantum amplitude (Feynman & Hibbs, 1965; Taylor, 2003; Ogborn & Taylor, 2005).

# 2 Principle of least action (PLA)

Let us consider two ways of how classical mechanics explains the motion of a falling apple: Newton's laws and Hamilton's principle of least action.[5]

**Newton's laws of motion**

Firstly Newton said: *Give me the apple's initial position and its velocity or two very nearby apple's positions*. Then Newton answered the question: *What is the position of the apple at the next instant if there is Earth's gravity or some force*. Newton postulated the first law of motion or the principle of inertia. If there were no acting forces, the apple would possess a mysterious internal tendency to continue in motion with the same velocity along a straight line. The second law of motion postulated that Earth's gravity or some force causes motion in the direction of the applied force. In other words, if the apple "perceives" at a distance the effect of the force, the apple is accelerated or changes its own the velocity. Thus, the path of the apple's actual motion is the result of the combining or summation of two tendencies or "effects": the apple's inertial motion and the motion due to the force acting on it. Finally, we obtain a differential equation to calculate all positions of the apple.

**Hamilton's principle of least action**

Hamilton said: *Give me both the apple's initial and final events (positions and times) in advance*. Then Hamilton answered the question: *Which path is followed by the falling apple between the initial and final events if the apple has potential energy*. According to Hamilton's principle, this is the path having the least action. The action is the difference between kinetic energy and potential energy integrated over time. This difference is called the Lagrangian and appears in Lagrange's equations of motion. As Feynman wrote:

> *"In other words, the laws of Newton could be stated not in the form F=ma but in the form: the average kinetic energy less the average potential energy is as little as possible for the path of an object going from one point to another"* (Feynman, 1964, p. 19-2).

---

[5] In this section, I use the description of Feynman (1964, p. 19-2) and Hanc (2006). Hanc, moreover, made it graphically for three approaches: Newton's laws, Hamilton's and Maupertuis' principles of least action.

It means that the action reaches the minimum, compared with all possible paths from the initial to the final events. Now, we do not need to know how the apple works its way from one event to another; we need to know only the initial and the final positions of the apple and times. Then, we must find all possible paths (or possibilities of movement) from the initial to the final events and by the so-called Euler variational method choose the one with the minimal action. Only this path is observed as the actual one, and it exactly coincides with the path calculated by the Newtonian approach. However, in Hamilton's principle, we do not think about forces. We also do not need the fictitious inertial force as, in the absence of the potential field, the apple's path of the minimal action is a straight line with constant velocity.

In this paper, the term *principle of least action* (PLA) covers not only Hamilton's principle but all integral variational principles of physics. These are utilized by mathematicians within the discipline of the calculus of variations and by physicists within analytical mechanics. The main point of each principle is to postulate the abstract space for a set of possible events, paths or histories for the system. According to the PLA, the actual history differs from all possible histories, consistent with the given constraints, that its function (called the action) is stationary and takes an extremal value. The actual history is that along which the system moves from one event to another within a specified space interval in configuration space.

In the calculus of variations, it is said that the variation of action upon infinitesimal variation in the history is equal to zero. In Hamilton's form of the PLA, the action of the body along the actual path equals an integral of the difference between the average kinetic and potential energy of the body. In a general case, the action can be calculated through an integral of the state function of the system over history, time, n-dimensional volume, or four-dimensional space-time. In most cases, the action is a local or global minimum; however, it may be a local or global maximum. Moreover, for any systems, the differential equations of motion could be derived from the PLA.

The PLA is not restricted to mechanics. Historically, the PLA arose from the optical-mechanical analogy with Fermat's principle, in which the light moves along the path taking the minimal amount of time. The PLA is used in electromagnetism, statistical mechanics, special and general relativity. According to Taylor's (2003) expression, a stone moving with non-relativistic speed in the region of a small space-time curvature obeys nature's command: Follow the path of least action! The stone moving with any possible speed in curved space-time obeys nature's command: Follow the path of maximal aging (or maximal proper time)! Here, Taylor kept in mind the relativistic analogue of the PLA – the principle of maximal aging (Taylor & Wheeler, 2000). Taylor (2003) also proposed a scheme where the PLA, on the one hand, is a limiting case of the principle of maximal aging, on the other hand, a limiting case of the of Feynman path integral formalism where an electron obeys the nature`s command: Explore all paths! In other words, Newtonian mechanics becomes a limiting case and an approximation of general relativity and quantum mechanics.

## 3 PLA and Feynman path integral (FPI)

When Butterfield (2004b) considered the variational principles of analytical mechanics throughout the philosophy of classical mechanics, he recognized the apparent fact that the actual world is quantum, not classical. One of the best illustrations of this fact is a deep relationship between the PLA and the FPI or the sum-over-histories model (Feynman & Hibbs, 1965). Indeed, quantum electrodynamics and the majority of quantum field theories are connected with the FPI, which uses the same notion of the action that the PLA does. The FPI calculates probabilities by summing up over classical configurations of variables and assigning a phase to each configuration, which equals the action of that configuration.[6] It is assumed that a quantum system simultaneously takes an infinite set of all possible alternative paths or histories, which correspond to the boundary conditions. In our classical world, these possible paths or histories are mutually exclusive even though at the quantum level these possible histories coexist. We can state these possible histories as being in quantum superposition. If the possible histories are coherent or mutually consistent (the difference between their quantum phases is constant), they state that there is a coherent quantum superposition. The probability amplitude of each possible history has an equal magnitude and varying phase, which corresponds to the classical action. The coherent or consistent histories are united due to the rule of interference. The resulting history has a maximal probability, which is given by the square of the sum of the probability amplitudes. It can be observed as a single actual history. It is important that other possible histories (called virtual or imaginary) do not disappear. They continue to be the necessary parts of the superposition though these histories are not observed because their probabilities are too small.

---

[6] This formulation of quantum mechanics is mathematically equivalent to the Heisenberg matrix method and Schrödinger wave equation, but intuitively more understandable.

As I know, Feynman did not insist on some philosophical interpretation of the FPI and quantum electrodynamics. At the beginning, he used his diagrams, which were directly connected with the FPI, not only as calculation tools but also as theoretically motivated representations of physical processes occurring in space and time. Later, he held a more abstract opinion (Wüthrich, 2010). However, he always understood that these approaches yield the same results as Newtonian laws in the classical limit (Feynman et al., 1964). To understand his view, let us imagine a classical body, as well as a photon or electron, moving simultaneously along all possible histories or world lines between the initial and final events. Whereas the phase of the quantum amplitude is very high, a set of world lines (making a significant contribution to the probability of the classical body's detection) is reduced to a narrow bundle. At the limit, the bundle shrinks to the single world-line predicted by the classical Hamilton's form of the PLA (Taylor, 2003). Thus, what Newtonian physics treats as cause and effect (force producing acceleration), the quantum "many paths" view treats as a balance of the changes in phase produced by the changes in kinetic and potential energy (Ogborn & Taylor, 2005). Thus, classical mechanics becomes a short-wave approximation of quantum mechanics, and the action acquires the meaning of the phase of quantum amplitude. The PLA of classical mechanics can be derived from the sum-over-histories model of quantum mechanics as a limit on a large scale. At the same time, the PLA is the limit of general relativity for low speeds and weak gravity (Taylor, 2003).

In quantum physics, it is generally accepted that possible paths or alternative virtual quantum histories in the FPI are merely formal mathematical tools for calculation, and it cannot be interpreted as implying that a quantum system actually follows one of the histories over which the FPI is computed. However, in recent years, there has been growing interest in a view of these possible histories having a certain degree of reality (Sharlow, 2007; Valente, 2011; Kent, 2013; Wallden, 2013; Wharton, 2013). In this paper, I do not consider any metaphysical issues of the FPI, even though it would certainly be an exciting subject. My aim is more modest – to make a metaphysical analysis of the PLA using FPI's model of the summation of all possible histories, along which the quantum object moves simultaneously.

# 4 Metaphysical issues of PLA

I do not have such an ambitious goal to cover all of the issues of the PLA or to criticise all the other literature that has tried to address them. I merely intend to present a more general evaluation of certain issues connected with causality, necessity, and a notion of possibility.

## 4.1 Causality and PLA

In the PLA, it appears that a physical system "foresees" in advance which path (of all possible paths for motion) will minimise an action. Feynman (1964) formulated this question in an original way:

*"Is it true that the particle doesn't just "take the right path" but that it looks at all the other possible trajectories? ... The miracle of it all is, of course, that it does just that. ... It isn't that a particle takes the path of least action but that it smells all the paths in the neighborhood and chooses the one that has the least action"* (Feynman, 1964, p. 19-9).

Although the various "as if" metaphors do not help us account for this old metaphysical issue of the PLA, we are compelled to do it. We undoubtedly cannot claim that natural objects "foresee", "smell", "calculate", or "choose" certain histories, especially the path of minimal action. At the same time, it appears that the observed events along the actual path are somehow connected with the future actual event. Indeed, why do physical systems behave in such way that one of their characteristics of actual motion takes an extremal value? In the history of science, there have been three opinions regarding the philosophical reasoning of the action's minimum: the perfection of God (the theological view), the economy of nature (the teleological view), and the economy of the human mind (the instrumental view).

After Leibniz, Maupertuis, and Euler, the teleological view of the PLA or the phantom of a final cause is considered too metaphysical and mystical (Goldstine, 1980; Lanczos, 1986). Hamilton disclaimed the economy of nature — referring to the action in his principle is a local or global minimum, but sometimes presenting as a maximum.

D'Alembert, Lagrange, Hertz, Jacobi, and other creators of the variational principles were sure that there are not any ontological foundations of the parsimony of nature. They considered the PLA to be only a figurative scientific model (Goldstine, 1980; Panza, 2003). Mach postulated the principle of the economy of thought, which in particular required preferring most economical, simple, and practical description of phenomena from all possible descriptions (Mach, 1907). He argued that the variational principles of mechanics are no more than other mathematical formulations of Newtonian laws and that they do not contain anything new (Stöltzner, 2003). Following Mach, Born emphasized that extremal descriptions talk not about properties of nature but about our aspiration for the economy of thinking (Born, 1963).

From the perspective of the absolute majority of modern physicists, the PLA is nothing but an equivalent method of mathematical description, such as differential equations. According to Yourgrau and Mandelstam (1968, p. 178f), the variational principles are closer to derived mathematical-physical theorems than to the fundamental laws. There are three arguments in favour of this point of view; however, each of them has certain weak aspects.

The first argument is that we can derive the PLA from the differential equations of motion. Yet, at the same time, the equations of motion are also derivable from just the PLA and the Lagrangian. It means that "we have no prima facie reason to think the equations of motion are fundamental" (Thebault & Smart, 2013). Especially if we take into account the fact that the PLA and the calculus of variation are more general than the differential equations. However, as Katzav (2004) pointed out, the mere fact that the PLA can be deduced from other equations does not point that there is such an explanation because deduction and explanation are not the same.

There are different models of scientific explanation (Salmon, 1989). According to the Deductive-Nomological model, to explain the phenomenon means to derive it from given initial conditions and at least one law. One of the conditions for such explanation is that the statements constituting the law must be true. Consequently, we have to be sure that the laws based on the differential equations are the true statements. The proponents of scientific realism believe that if a theory is in good agreement with the experiments, such theory postulates on things and events that exist and occur in fact. Therefore, we must believe that the theory is true. However, there are a number of serious objections in opposition to this view (Dewitt, 2013): (a) in the past, a good many theories turned out to be false; (b) experiments are always limited; (c) in the heart of every theory there are certain models with many idealizations and abstractions, and thus the laws of the theory are valid only for its models. Therefore, we cannot be sure that the differential laws are the true statements.

A further issue is that the same phenomenon can be predicted by several mathematically equivalent theories, each with its own explanations. To illustrate this, Feynman (1985, pp. 50-53) stated the law of gravitation in three different ways, "all of which are exactly equivalent but sound completely different". These are: Newton's law, the local field method, and the minimum principle (he meant the PLA). Feynman emphasised that they are equivalent scientifically, but "philosophically you like them or not like them", and "psychologically they are completely unequivalent when you are trying to guess new laws."

Indeed, the common philosophical and psychological preferences of scientists arouse two additional arguments in favour of the instrumental view of the PLA. The second argument refers to the general principle of causality based on the grounded belief that causal influences cannot travel backwards in time. The disadvantage of this argument is that it is directed only against the PLA. Although, most of the physical laws are time symmetric. It seems there are three options. First, the physical laws are true and causal processes are intrinsically symmetric in nature, therefore forward causation and backward causation are always subjective. Second, the causal asymmetry is objective, thus we should reconsider the classical physical laws. Third, "there would be nothing conceptually, nor physically, that could distinguish backward causal processes from forward causal processes" (Faye, 1997, p. 262). The last option is coordinate with Planck's (1958) view of final causes in the PLA as a merely alternative form, but, in fact, an equal point of view of ordinary causality of efficient causes. It means that the PLA and other variational principles of physics do not show any benefits to either determinism or teleology (Whitrow, 1980).

The third argument in favour of the instrumental view of the PLA refers to the principle of locality, is that cause and effect are connected across space and time. According to the relativity theory, causal influences cannot travel faster than the speed of light, and cause and its effect are separated by a time-like interval. Among other things, this means that an object cannot be simultaneously in the different places or move at the same time along various trajectories. However, in recent decades, the principle of locality subsequently is being questioned by the quantum experiments (Henson, 2005; Pietsch, 2012).

In Section 3, I have mentioned the quantum formalism of the FPI that uses the same "all at once" principle. The FPI considers that the quantum system as if simultaneously takes an infinite set of all possible alternative histories from one event to another. The question is whether this coincidence is just formal and accidental or based on an unknown, metaphysically necessary law.

**4.2 Necessity of PLA**

Despite the significant place of the PLA among the laws of motion, philosophers have not sufficiently examined whether this principle is truly necessary or not. Many philosophers, beginning with David Hume, have argued that the laws of nature are metaphysically contingent truths. According to the Humean view (or the Regularity Theory), these laws are mere regularities, expressed by the universal quantifications that form part of the best system of law-statements (Lewis, 1986).

Other philosophers have held that all (Shoemaker, 1998; Bird, 2005) or some (Ellis, 2001) laws of nature are necessary truths, not contingent; and that physical possibility is equivalent to metaphysical possibility.[7] Some philosophers argue that the laws might differ in the kinds of their necessity (Fine, 2002). Others have suggested that not all laws of nature are necessary in the same way; that certain laws of nature might be more general than others because they have different degrees of necessity. Lange (2007), for instance, takes symmetry principles as meta-laws governing ordinary laws. Consequently, according to the non-Humean view (or the Necessitarian Theory), there are necessary connections between events, and we must reject the theory of Humean Supervenience and implement a new kind of realism in philosophical analysis (Ellis, 2001).

Both opposing views raise some questions. If the Humean view is correct, are the differential laws of motion (e.g., Newtonian laws) and variational laws (e.g., the PLA) metaphysically accidental to an equal degree? If so, then why is the PLA being considered one of the most fundamental laws, from which all other laws of motion can be derived? If the non-Humean view is correct, could we say that the PLA is a metaphysically more fundamental law governing the physical phenomena of the world? If the latter is true, we face three consequent questions.

---

[7] Within the metaphysics of modality, there is a debate about the relationship between physical and metaphysical possibility. It is accepted that the physical area of possible events is narrower than the metaphysical ones. The metaphysically possible event is possible by virtue of its own essence or true in one of the metaphysically possible worlds. A metaphysical necessity refers to essence or truth in all metaphysically possible worlds. Something is considered as physically possible if it is permitted by the laws of physics. Respectively, a physical necessity directly follows from these laws (Vaidya, 2015).

(1) How can other laws of motion (e.g., Newtonian laws) be mathematical and logical consequences of the PLA?

(2) How does the metaphysical necessity of the PLA involve contingency of the classical system's possible histories and uncertainty of the quantum system's probability amplitudes?

(3) What is the source of the metaphysical necessity of the PLA?

In Section 9, to answer these questions, I broaden the notion of the physical necessity and suggest two new notions: the laws with a limited physical necessity and the laws with a general physical necessity.

### 4.3 Possibilities and PLA

Butterfield (2004a) did not agree with philosophers, who said that the virtual displacements or variations mentioned in the variational principles have nothing to do with possibilities of the sort discussed in modal metaphysics. He was even convinced that "mechanics is up to its ears in modality, of some kind or kinds." However, other philosophers do not consider the issue of the reality of the possible (virtual) paths or histories in the PLA and other variational principles. Moreover, from the point of view of physics, there is no problem as the notion of "possible history" is no more than a heuristic and mathematical tool for writing the laws of motion.

From the perspectives of metaphysics, there is a weird state of affairs. On the one hand, all possible histories are logically possible and "exist" only in our minds. On the other hand, we might consider and calculate the possible histories in which the physical system could have evolved in reality. A philosopher faces an issue. What if the possible histories in the PLA possess some grade of reality? What if they take place in the semi-real space of a possible event? If this assumption is true, we face another set of questions:

(1) How does a multitude of possible histories turn into the actual history, or why do only some of the possible histories become actual? Does this transformation occur accidentally or by law? If this happens by law, how is the selection made?

(2) What happens to the possible histories that never become actual?

(3) Could we describe the possible histories in the PLA by using the metaphysical theories of possible worlds and possible objects?

In discussing a modality in analytical mechanics, Butterfield (2004a) considered only David Lewis' theory of counterfactuals. I believe that the wide metaphysical debate about the nature of possible histories is the most promising way to explain the metaphysical issues of the PLA. In Section 5, I involve certain modal concepts in the discussion regarding the ontological status of PLA's possible histories. Then I use them to support my modal interpretation of this principle.

## 5 PLA, possible objects, possible histories, and possible worlds

There is a wide-spread opinion that states that any object is an actual object. From this follows the concept that non-actual possible objects are nothing. There is another conservative view that any object is an existing object. However, from a metaphysical perspective, these statements are not obvious. Moreover, the issue of the reality of possible objects (so-called *possibilia*) is one of the most difficult challenges of metaphysics.

First of all, the analytical philosophers have been paying special attention to the correlation between the *being*, *existence* and *essence* of possible objects and possible states of affairs in different possible worlds (Adams, 1974; Lewis, 1986; Fine, 1994; Armstrong, 1997; Divers, 2002). Some theories concerning possible objects do not invoke the possible worlds. For instance, *essentialism* is a doctrine where objects have essential properties in terms of an entity's *de Re*[8] modal properties (Fine, 1994), and the so-called *Meinongian approach* constructs a general theory of objects other than ordinary concrete existing objects (Zalta, 2006).

Of course, the possible histories in the PLA do not equal the possible objects. Moreover, we can imagine two types of possible histories – for both possible and actual objects.

(a) An actual object is defined by its actual states in actual space or actual events in actual space-time (we can call it the actual world). The set of such consecutive actual events forms an actual history. Accordingly, a possible object is defined by its possible events in the same or another space-time of possible events (we can call it the possible world). Therefore, the set of the consecutive possible events could be considered the *possible history of the possible object.*

---

[8] According to modal logic, if a statement is true in all possible worlds, then it is necessary. A statement that is true in some possible worlds is possible. To emphasise the difference between modal logic and metaphysics of modality, the philosophers often divide modalities into two kinds: *de Dicto* and *de Re*. The second kind consists of the modalities that are inherent in things and phenomena, regardless of our language.

(b) An actual object possesses its possible states in actual space or actual events in actual space-time. The set of such consecutive possible events could be considered the *possible history of the actual object*. It means that the same actual event could be reached by many possible histories. At the same time, many other possible histories may start from one actual event. The main restriction is that the possible histories of the actual object must be consistent with the physical laws of the actual world.

Before suggesting my approach regarding the nature of the possible histories in the PLA, I briefly review several metaphysical concepts of the possible objects (possibilia) and possible worlds.

The simplest way to address the issue of the nature of the possible objects (possibilia) is to use the metaphysical notions of *the being* and *existence*. However, we always need to specify where something is or exists. For instance, the radical possibilist or modal realist (Lewis, 1986) states that the possible objects and possible events have being and exist no less than the actual ones in an infinite number of possible worlds. For the classical possibilist (Russell, 1903, §427), every existing object is (ontologically) in our world, but some of them (possibilia) could only have existed there. According to the most commonsensical position, such objects are not actual and do not actually exist, but they have a certain grade of being. The actualist denies any reality of possibilia, which are mind-involving and exist only as names, fictions, "ersatz" linguistic, or theoretical constructions (Adams, 1974; Armstrong, 2004). Thus, everything that is, exists as an actual thing, and physical existence equals being. Some of the actualists (Plantinga, 1974) invoke unactualised individual essences. They stated that every object has an individual essence that is independent of the object possessing it, whether the object is actual or non-actual.

Bird (2006), in a discussion with Armstrong (1997) about potency as an essentially dispositional property, demonstrated that a key problem is that ontology uses a number of terms to describe perhaps different and unequal kinds or degrees of being. One can say of something that it *is*, that it *exists*, that it *is real*, and that it *is actual*. One says that the merely possible objects are not real; they do not exist because possibilia (such as unrealised manifestations of potencies) are a violation of naturalism. According to Bird, a source of this mistake lies in a picture dominated by modal realism (possibilia cannot exist in the actual world but can exist in other possible worlds). Thus, we are faced with a dilemma: if we accept modal properties we accept other possible worlds. However, this seems to conflict with causal naturalism. Bird's solution is to reject modal realism. Possibilia are not things that exist (if at all) in other worlds, but not in our world. Instead, possibilia are things that have being in our world but do not exist. Thus, unrealised manifestations of possibilities are part of the world just as much as manifestations that are realised.

Let us pass from the possible objects to the possible histories. Every object can be considered a unit of two aspects: static (an event) and dynamic (a set of consecutive events or a history). Since the PLA describes the movement of the actual objects, I consider only the possible histories that comprise the set of the consecutive possible events in the actual space-time of our world.

If we apply the above mentioned above metaphysical theories to the PLA, we could say that:

(in modal realism) the possible history *exists* in possible worlds, but not in our actual world; in the actual world, we can observe only the history with the minimal action;

(in possibilism) the possible history has *some grade of being* in our world but does *not exist*; the observed history with the minimal action has the full being and consequently exists in the actual world;

(in actualism) the possible history *is* a name or fiction and does *not exist* in our world, even though it can have some individual *essence*; we can observe the actual history with the minimal action because it is not a fiction and does not depend on our minds;

(in dispositional essentialism) the possible history *is* in our world as unrealised manifestations of possibilities, but the possible history does *not exist* there. The observed history with the minimal action is the realised manifestation of one of possible histories and thus, it exists in the actual world. In Section 7, I consider other options of applying dispositional essentialism to the PLA.

**Leibniz**

Although most of the theories of possible worlds are based on the ideas of Leibniz, his metaphysical system is significantly different from almost all of modern modal theories and deserves a separate study.⁹ For Leibniz, actuality is something that expresses the existence but potentiality expresses only the essence. Since Leibniz was a scientist no less than a philosopher, he was dissatisfied with the overly abstract Aristotelian model of the implementation of the potentiality (*dynamis*) through activity (*energeia*) to the actuality (*entelechia*). Leibniz tried to imagine how this metaphysical process manifested itself in physical processes.

Leibniz's theory of the striving possibles (1989, pp. 149-155), distinguished between essence (the nature of a thing) and existence. He postulated that the principle of governing essences is that of possibility or non-contradiction. He suggested that each essence (each possible thing) tends of itself towards existence, but the one that will actually exist is that which has the greatest perfection or degree of the essence or the greatest number of possibilities at the same time. The more perfection, the more existence. According to Leibniz, the things are incompatible with the other things; therefore, some possible things do not achieve their actualisation. From the collision of all possibilities, only those things that contain the greatest number of possibilities will be actualised. In other words, "the possibles vie with one another for existence by combining forces with as many other essences as they are mutually compatible with" (Blumenfeld, 1973). Thus, the world arises in which the largest part of the possible things is actualised. Leibniz gave physical examples of such things: a straight line among all lines, a right angle among all angles, and a circle or a sphere among all figures as the most capacious ones.

If we expand the Leibnizian doctrine of the striving possibles and apply it to the PLA[10], we could say that:

(in Leibnizian) every possible event or possible history has its *essence* and tends towards *existence* in our actual world. Among the infinite set of the possible histories, only the history with the minimal action can *exist* as actual because it has the highest degree of the essence and combines the greatest number of possibilities at the same time.

It seems that despite many differences, the views of the possible histories based on the contemporary metaphysics of modality and the Leibnizian theory have some resemblances and analogies. Therefore, I postulate that there are sufficient reasons to suggest a new approach regarding the nature of the possible histories in the PLA. Let us start from a two-level model of being or two realms of our world.

---

[9] Leibniz argued two concepts of possibilities, which had long been rejected by most philosophers. The first was that God has an infinite number of possible worlds. We can be aware of all them because, according to Leibniz, being is inherent in everything that can be thought, but not everything obtains being. Thus, by the will of God, only the most perfect world is actualised. The second was the doctrine of the striving essences or possibles. The second, however, seems plainly inconsistent with the first. See the discussion about this (Blumenfeld, 1973; Shields, 1986).

[10] Some advocates of the PLA, among them Planck, considered Leibniz as the discoverer of the PLA (Planck, 1958, s. 24).

At the first level, there are the possible events and histories of the actual objects. The possible events and histories have essences but do not actually exist. We can call this level a possible modality of being or a possible realm of the world.

At the second level, there are only actualised events and histories of the actual objects. The event and histories have both essences and existence. Unlike the possible histories, the actual ones possess more dispositions towards existence or a higher degree of the essence. We can call this level an actual modality of being or an actual realm of the world.

## 6 Modal interpretation of PLA

To provide a positive metaphysical solution for the issues of the PLA mentioned in Section 4, I consider a hypothesis concerning the nature of the possible histories in this principle. The hypothesis is based on the two metaphysical models: modality and combination. These models together form a modal interpretation of the PLA.

### 6.1 The modality model

The modality model is a statement of the two-level modality of the physical system's histories. It is based on various modal approaches to the reality of possibilia in the contemporary metaphysics of modality and Leibniz's theory of the striving possibles, which are described in Section 5.

In the modality model, I do not involve the possible worlds. The possible worlds are important metaphysical notions, but even if they exist, this fact does not matter for an analysis of the PLA. Moreover, in metaphysics, Leibniz's idea of many possible worlds became an obstacle to the development of his other ideas of possibilities. Philosophers often connect the possibilia with possible worlds because of the authority of Lewis.[11] In my opinion, the possible objects, possible events, possible histories, and Lewisian worlds are not the same; thus, we cannot use Lewis' arguments. Here, I agree with Bird's (2006) opinion given in Section 5 that a source of mistake lies in a picture dominated by modal realism (possibilia cannot exist in the actual world but can exist in other possible worlds). Bird's solution is to reject modal realism. Possibilia are not things that exist (if at all) in other worlds, but not in this one; instead, they are things that have being in our world but do not exist. According to a further possible objection, in the realm of classical physics, the possible worlds are set up to determine the actual world, while, in the quantum realm, possible worlds contribute with a certain probability. Here, the quantum object's possible histories with a certain probability are confused with the possible worlds.

---

[11] Butterfield (2003) discussing various modal involvements of the variational principles tied them to David Lewis' work on modality, especially to his work on counterfactuals (Lewis, 1973). Butterfield even took an instantaneous state as the analogue of a Lewisian world.

According to the modality model, the possible histories, mentioned in the PLA, have essences in the possible modality but do not actually exist in the actual modality. The actual or actualised histories have existence in the actual modality. Now, I specify what the essence and existence mean for the possible histories of the PLA. The absence of existence means just the absence of physical observation and interaction in the actual modality. At the same time, the possibility of the histories means their non-contradiction of the classical physical laws of the actual realm of our world. Since the possible events and histories have essences, they can occur simultaneously in different possible space-times of the possible realm of our world. The properties of these possible space-times are the subject of a special investigation. The actual history is naturally consistent with the physical laws of our world and occurs in the only actual space-time.

In Section 5, I applied the Leibnizian doctrine of the striving possibles to the PLA. I connected the essence of possible histories with the tendency towards existence. However, Leibniz meant that all possible things have the essence. This can also mean that each possible object, each possible state, and each possible event tend towards existence. For an actual physical system, it means that the system tends towards existence in all possible ways or moves from each initial actual event along all possible histories. It is important to emphasise that this is not like the way in which a pendulum will eventually come to a stable equilibrium due to air resistance, because the pendulum passes all possible states sequentially one by one. In my modality model, these possible movements occur simultaneously in the possible modality.

Such a picture contradicts the classical laws because the physical system's possible histories are mutually exclusive or not compossible in four-dimensional space-time. Unexpectedly, quantum mechanics can help here. In Section 3, I have mentioned the deep relationship between the PLA and the FPI. To calculate the probabilities of the quantum particle's history, the FPI supposes that the quantum system simultaneously takes an infinite set of all possible alternative histories corresponding to the boundary conditions. At the quantum level, these histories coexist, and physicists say that the possible histories are in quantum superposition.

Using an analogy with the FPI, the modality model claims that all of the physical system's possible histories *are* jointly in the possible modality of being or possible realm of the world. The possibility of the histories means their non-contradiction, both in the classical and quantum physical laws. In other words, the modality model replaces the classical representation of a system's motion along a single actual trajectory by a representation of simultaneous motions along an infinite set of possible histories. The possible motions occur simultaneously in the possible realm of our world.

**6.2 The combination model**

Another part of the modal interpretation of the PLA is the combination model, which is a statement of the combination or integration of all the physical system's possible histories. This model aims to explain of how the set of the possible histories in the possible modality turns into the actual history and why only some of the possible histories become actual, as well as how this selection occurs. I will call initially upon several parallels from physics and then the metaphysical ideas of Leibniz.

The first parallel comes from classical physics. Let us remember the physical models where certain actual movements are considered to be the combination and summation of a set of the possible or virtual movements. Section 2 has illustrated that in the Newtonian approach, the actual motion is the result of the combination or summation of two possible motions: the inertial motion and the motion due to the acting force. In optics, Huygens' principle states that all points of a wave front of light may be regarded as new sources of virtual secondary waves that expand in every direction. The sum of these secondary waves (or a surface tangent to them) constitutes the new actual wave front at any subsequent time.

Another kind of the combination is used in the integral variational principles. A certain functional of a system (not just the action) is stationary and takes an extremal value for the actual process or history among all alternative possible processes. The functional is defined by the integral of a certain expression (the Lagrangian or Lagrangian density), and can be calculated over the path, time, n-dimensional volume or four-dimensional space-time. Unlike the integration variational principles, the differential variational principles use the summation that set equal to zero.

In quantum physics, the physicists also involve the imaginary model of the summation of many trajectories or histories. Heisenberg (1962) believed that an observed trajectory of a single particle resulted in one of the possible trajectories transforming into the actual trajectory. In contrast to Heisenberg, Schrödinger (1965) explained the actual trajectory by a set or a field of all possible trajectories. Instead of the logical opposition between an "either–or" in point mechanics, he proposed using a "both–and" in wave mechanics. According to his wave mechanics, "the infinite array of possible point paths would be merely fictitious, none of them would have the prerogative over the others of being that really travelled in an individual case" and all of them are equally real. Moreover, we cannot manage to make do with "such old, familiar, and seemingly indispensable terms as "real" or "only possible"; we are never in a position to say what really is or what really happens, but we can only say what will be observed in any concrete individual case." In another paper, Schrödinger (1952) underlined that instead of the implementation of only one possible entangled state, all are summed up. It occurs due to the resonance or interference of the waves.

As discussed in Section 3, according to the FPI, the coherent histories are united due to the rule of interference or the summation of quantum phases. Each quantum phase corresponds to the classical action, and (in the general case) the resultant actual history obtains a maximal probability and minimal action. The actual history can be obtained as the limit of a narrow bundle of the possible histories significantly contributing to the quantum amplitude. It is important to emphasise that other possible histories do not disappear and continue to be the necessary parts of the quantum superposition, though they are not observed due to their incredibly small contributions to the probability amplitude.

In the modal interpretation of the PLA, the combination model of the physical system's possible histories is based on an analogy with the mathematical operations of summation or integration. With the analogy of quantum superposition, we can deal with the metaphysical superposition of the coherent set of the possible histories in the possible modality. Due to the combination of all the system's possible histories, only the resultant history obtains existence in the actual modality, and it is the only one that becomes observable in the four-dimensional space-time of the actual realm of our world. Hence, the actual history is the necessary combination of all possible histories although, each possible history is accidental and has its own probability.

According to the analogy between the modal interpretation of the PLA and Leibniz's scheme of the striving possibles, every possible history has the essence and like all possibles or "everything that expresses essence or possible reality, strive with equal right for existence in proportion to the amount of essence" (Leibniz, 1989, p. 150). I postulate that the actions of the possible histories somehow correlate with the essence. Indeed, then Leibniz stated that "of the infinite combinations of possibilities and possible series, the one that exists is the one through which the most essence or possibility is brought into existence" (Ibid). Accordingly, in the modal interpretation of the PLA, of the infinite set of the possible histories, only the one with the minimal action can exist as actual because it has the highest degree of the essence and combines the greatest number of possibilities at the same time. It appears the more essence a possible history has, the less action there is. Nevertheless, it is not exactly so. As we know, in the integral variational principles, a certain system's functional (not just the action) is stationary and takes a minimal or maximal value for the actual process among all alternative possible processes. It means that the essence and the action are not exactly the same; the former is not a definition of the latter and vice versa. Rather the metaphysical interpretation of the action (and certain system's functionals) is one of the physical measures of the essence, which consists of the necessity of each possible history to be realised in actuality.

Let us continue the reasoning of Leibniz. It appears, in the possible realm of the actual world, an actual system uses the maximal number of the possibilities of motion in each subsequent actual event, thus it moves simultaneously along all possible histories from each point of space and each moment of time. A certain kind of a collision or even "competition" occurs between these possible histories. The result of such a "competition" has the maximal essence and is manifested in the actual existence as a unique history. Other possible histories do not disappear completely since they remain in the possible modality. They still have essences, but they are not compossible. Only mutually compatible essences constitute the actual world. Here, I follow Leibniz, who invoked the notion of compossibility, so that "the universe is only a certain collection of compossibles, and the actual universe is the collection of all existing possibles, that is to say, those which form the richest composite" (Messina & Rutherford, 2009).

To sum up, all possible motions occur simultaneously in the possible realm of our world. The system moves at the same time along all possible histories from each point of actual space and each moment of actual time. The combination model states that only the actual history between any two actual events has the highest degree of the essence and combines the maximal number of possible histories between these events at the same time. In moving along the actual history only, the system can take the maximal number of possible states. This principle works at each moment of time, as well as for a finite duration of time. All systems move simultaneously along all try every possible history. The continuous mutual play of the system's motion along all possible histories in the possible modality and the following combination of these motions create the system's actual events and actual history and as a result, the entire actual realm of our world.

From a physical view, it appears in the rule of the summation of the possible history's actions or certain system's functionals. Consequently the significance of the PLA, and perhaps other variational principles, lies not with the mystic economy of nature, but merely in the observable effect of the combination of the possibilities to move.

As we can observe, the modal interpretation of the PLA has some relations with the approaches of modal metaphysics (Section 5). It is certainly different from actualism, in which alternative, unrealised possible histories are fictions or theoretical constructions only. At the same time, it is close to Plantinga's (1974) view of an individual independent essence of the non-actual object. My hypothesis also closely resembles dispositional essentialism, as the possible histories can be treated as manifestations or potencies unrealised in the actual world.

## 7 PLA and dispositional essentialism

Recently, the relationship between the PLA and dispositional essentialism provoked rich discussions (Katzav, 2004, 2005; Ellis, 2005; Bird, 2007; Thébault & Smart, 2013). According to dispositional essentialism,[12] at least some fundamental properties have objective propensities or dispositional essences that nudge the outcomes one way or another. The world is, ultimately, merely something like a conglomerate of objects and irreducible dispositions. The dispositional properties are, unlike categorical properties, supposed to be properties that are not wholly manifest in the present; thus, they are the ultimate ontological units that explain events. Any object that possesses the dispositional essence of some potency is disposed to manifest the corresponding disposition under stimulus conditions, in any possible world (Shoemaker, 1984; Ellis, 2001; Bird, 2006, 2007).

---

[12] There are other attempts to consider the propensities and dispositions as something real, see, for example (Suárez, 2004, 2011).

Applying dispositional essentialism to the PLA (also see Section 5), we can translate it as follows: Each point in velocity-configuration space represents an instantaneous pattern of dispositions or dispositional and categorical property instantiations, and the history represents the actual evolution of the system through various states (Thébault & Smart, 2013).

Katzav (2004) argued that dispositional essentialism is not compatible with the ontological presuppositions of the PLA and, ultimately, dispositionalist ontology is not able to account for the metaphysical presuppositions of science. Katzav made the assumption that the PLA suggests that the action of any given physical system could have taken various values, and, thus, that any such system could have been correctly described by the different equations of motion. The PLA allows us to derive the equations of motion of a system by comparing the various quantities of the action that the system might have had rather than by appealing to the system's actual history, it does not offer a historical explanation for why the actual equations of motion are actual. The PLA requires that dispositions do supervene on nondispositional properties taken together with something like a law, namely whatever makes the PLA true.

In reply to Katzav, Ellis (2005) argued that only a sophisticated dispositionalist can accommodate the PLA and its metaphysical necessity. He proposed that how things are disposed to behave also depends on how the kinds of things and properties are placed in the natural kinds of hierarchies. Ellis claimed that the PLA is of the essence of the global kind in the category of objects or substances. Then every continuing object must be disposed to evolve in accordance with the PLA.

Thébault and Smart (2013) argued with Katzav and stated that dispositional essentialism is consistent with the PLA. One of the reasons is that there is only one metaphysically possible history in which the physical system could have evolved, but this still allows for there to be many logically possible histories. Despite all their arguments and objections, Thébault and Smart accepted that the dispositionalist "has no teleological metaphysical interpretation explaining the important and surely non-accidental PLA." Katzav, in his turn, left open the question of whether dispositionalism remains viable. He proposed that we might try to maintain dispositionalism, for example, by combining the instrumentalist view of the PLA with the realist view of the equations of motion.

I suggest reconciling the PLA with dispositional essentialism in another way based on the modal interpretation of the PLA. We might suppose that each object's possible history possesses its own disposition.[13] According to Bird (2006), possibilia (such as unrealised manifestations of potencies) are things that have being in our world but do not exist. Thus, unrealised manifestations of possibilities are part of the world just as much as realised manifestations. In Section 5, I proposed that if we apply such a view of possibilia to the PLA, we could say that the possible histories have essences and being in the possible modality as unrealised manifestations of possibilities, but the possible histories do *not actually exist* in the actual modality. Thus, the observed history with an extremal action is the realised manifestation of one of the possible histories and exists in the actual world.

According to the modality model of the PLA (Section 6), the actual or actualised histories have existence in our world due to their higher degree of the essence. The dispositions of actualised histories differ by degrees of necessity in being manifested in the actual modality, and the degree of necessity can be measured by the value of the action. It seems that the dispositions also "compete" with one another. The result of this "competition" with the maximal disposition, maximal degree of necessity, and minimal action is realised in the actual history. The "competition" of the dispositions occurs simultaneously in the possible realm of our world. Other dispositions remain unrealised.

## 8 Modal interpretation of PLA and causality

One of the metaphysical issues of the PLA relates to causality (Section 4.1). It is as if a physical system "foresees" in advance which history (of all possible histories of motion) will minimise an action. The system seems to "choose" the actual history along which an action is less than of along other histories. It seems as if the system's final state determines the history that the system takes to reach that state, or causal influence travels backwards in time.

---

[13] Popper (1990), for example, had seen a world of propensities, as an unfolding process of realising possibilities and of unfolding new possibilities; and the propensities or dispositions that have not realised themselves, have their own reality. Each of these propensities has an objective measure, which can be associated with probability. Concerning quantum objects, the Popper's theses were that the propensities are the relational properties of the quantum entities in experimental set-ups; the quantum wave function, or state, is a description of a propensity wave over the outcomes of an experimental set-up.

Let us examine how the modal interpretation of the PLA (Section 6) can change the view of causality in the PLA. We replaced the classical representation of a system's motion along a single actual history by a representation of simultaneous motions along an infinite set of possible histories. All possible motions occur simultaneously in the possible realm of our world. The system moves at the same time along all possible histories from each point of actual space and each moment of actual time. Only the actual history between any two actual events has the highest degree of the essence and combines the maximal number of possible histories between these events at the same time. This principle works at each moment of time, as well as for a finite duration of time.

It is no longer necessary to suppose as if the system "knows", in advance, which of its histories possesses the minimal action and thus will be the actual history. The system does not need to "choose" anything. Rather, it merely uses the maximal number of possibilities of motion in each subsequent actual event. To achieve this aim, the system it is sufficient to move simultaneously along all possible histories. All systems do the same in the possible modality; they simply try every possible history. The continuous mutual play of the system's attempts in the possible modality and the following combination of these attempts create the system's actual events and actual history and as a result, the entire actual realm of our world. With regard to the PLA, the system does not need to "calculate" the value of the action or anything else, the rule of the summation of the history's actions does so, since the action is merely the physical measures of history's essence.

On the one hand, this modal approach does not need the backward causation because some possible movements can occur forward in time. On the other hand, it does not prohibit other possible movements from occurring backward in time. It seems, in the possible modality, the certain direction of time does not matter, and all possible directions of time have equal rights. In the actual modality, we always observe the forward causation as the combination of the possible movements as forward as well backward in time. In other words, the physical laws are time symmetric due to the absence of the selected direction of time in the fundamental possible realm of our world.

As mentioned in Section 4.1, according to the relativity theory, causal influences cannot travel faster than the speed of light, and cause and its effect are separated by a time-like interval. This also means that an object cannot move simultaneously along various trajectories, which are mutually exclusive in the actual space-time. However, this principle of locality refers only to the actual space-time. In the possible realm of our world, all speeds of light and all time-like intervals are possible. According to the modal interpretation of the PLA, the possible histories occur simultaneously in different possible space-times, and after the combination of these histories, one of them obtains existence in the four-dimensional space-time of the actual realm of our world, which obeys the principle of locality.

The modal approach to the analytical mechanics can help us with the fundamental metaphysical challenge of the way in which the universe actually works. We commonly predict a future state of a system if we know the initial state and the differential equation of the dynamic law. From this, we sometimes conclude that the universe makes the same and as if it "calculates" its own states successively, one by one. In other words, we believe that in the metaphysical law, the past causes the future. However, more often, we can predict a future state of a system by using the PLA or other principles of analytical mechanics. Wharton (2015) posed two unexpected questions: what if the universe works in a different way and does not follow Newton's rule but follows Hamilton's and Lagrange's? What if the universe does not "consider" its own states and histories successively but all at once? From the perspective of the modal approach to the analytical mechanics, it is precisely such a picture that can be plausible since any actual state is the total result of all possible histories connecting the past and the future.

## 9 Between Humean and non-Humean concepts

I believe that the modal interpretation of the PLA can both address the metaphysical issue of the necessity of this principle (considered in Section 4.2) and provide a new view of the laws of physics. The Humean recognises only one actual world, and that the laws of this world are descriptions of regularities exhibited by the events in the actual history of our universe. Modalities, like possibility, necessity, and counterfactual statements are introduced as conceptual tools that enable us to deal theoretically with the actual world; they do not have an independent life of their own (Dieks, 2010).

At first glance, the PLA is the same law as all of the others; and the Humean concept of the laws of nature, without amendment or discomfort, is happily committed to the PLA, being the most fundamental law of nature (Thébault & Smart, 2013). These authors have argued that the PLA is a law in virtue of the history (that the physical system follows) is that which extremises action. Though they have agreed that, even with regard to the Humean view, the PLA is the most fundamental law, from which all other laws of nature can be derived. Here, "the fundamental law" does not refer to some metaphysical essence; rather, it refers to the PLA having a more explanatory power. However, this seems to be a weak argument. The Humean view does explain neither why the most fundamental law appeals to the strange notions of the possible event and possible histories nor how these differ from the actual ones.

Thébault and Smart (2013) claimed that, from the Humean perspective, the PLA has the explanatory value. Here, they partly agree with Katzav (2004), though according to them, the real explanatory role is played by whatever makes the PLA non-accidental. Katzav showed that the explanatory force of the PLA is founded on the fact that certain quantities are extremal. It seems to imply that, if the history is actual, the actuality is not an accident; moreover, that something is not an accident enables appealing to it in explanations. Unfortunately, Katzav did not provide any positive metaphysical account of the PLA.

The arguments for the non-Humean view are insufficient as well. In Section 4.2, I proposed that if the non-Humean view is correct and the PLA is metaphysically necessary truth, we face three consequent questions.

(1) How can other laws of motion (e.g., Newtonian laws) be mathematical and logical consequences of the PLA?

(2) How does the metaphysical necessity of the PLA involve contingency of the classical system's possible histories and the uncertainty of the quantum system's probability amplitudes?

(3) What is the source of the metaphysical necessity for the PLA?

Now, I outline a hypothesis whose answers lie between the Humean and non-Humean concepts of the nature of the PLA. To explain how it is possible, I will suggest two new notions: the laws with a limited physical necessity and the laws with a general physical necessity.

First off, let us return to the idea of the two-level modality considered at the end of Section 5. The first level is the possible modality of being or possible realm of the world. The second level is the actual modality or actual realm of the world. According to the modal interpretation of the PLA (Section 6), the possible histories have essences in the possible modality but do not actually exist in the actual one. The actual or actualised histories have existence because, unlike the possible histories, they have more dispositions towards existence or a higher degree of the essence.

This model means that, in the actual modality, the PLA is a mere regularity since the action's minimum is not a necessary reason for a history to be actual. The action could also be maximal or take a stationary value. Thus, the PLA seems to be a metaphysically contingent truth. However, in the actual modality, the PLA and other variational principles are universal tools of study regarding how various physical systems move, and they cannot be accidental. I agree with Thébault and Smart (2013) that the real explanatory role is played by something that makes the PLA non-accidental. I propose that the unknown something that makes the PLA non-accidental is necessary and lies in the possible realm of the world. The PLA is the fundamental law governing some kinds of physical motion, and the Newtonian laws are its mathematical consequences; however, in the possible modality, this is not the case.

Now, let us divide all physical laws of motion into those that are necessary only for certain kinds of physical systems and those that are necessary for any of the physical systems in our universe. Let us call the former *the laws with a limited physical necessity* (LPN-laws) and the latter *the laws with a general physical necessity* (GPN-laws). The LPN-laws involve only the actual realm, and only actual motions can be described by these laws. The LPN-laws lack any metaphysical grounds; thus, they are relative and metaphysically contingent truths. At the same time, the LPN-laws follow the GPN-laws, which are based on certain *laws with a metaphysical necessity* (MN-laws) and also involve the possible motions in the possible realm of our world.

It appears that the PLA and other variational principles are the GPN-laws.[14] On the one hand, PLA's necessity is wider than that of the Newtonian and other differential laws, which definitely work only for actual objects. On the other hand, according to its modal interpretation, the PLA is the necessary consequence of two MN-laws. The first MN-law corresponds to the modality model of the PLA (Section 6) where all systems, in each actual state, tend to actualise the maximal number of their possibilities for motion. This law calls for the actual objects to move simultaneously along all possible histories in the possible realm of our world. The second MN-law corresponds to the combination model of the PLA, where due to the combination of all the system's possible histories, only the resultant history obtains existence in the actual modality, and it is the only one that becomes observable in the four-dimensional space-time of the actual realm of our world. Thus, the second MN-law calls for the actual history to be the sum of all possible histories.

---

[14] The GPN-laws, in certain sense, are similar to meta-laws of Lange (2007). He argued that symmetry principles are meta-laws governing ordinary or "first-order" laws in a manner analogous to the way in which those laws govern ordinary facts. I agree with Lange that the symmetry principles qualitatively differ from other physical laws. It appears that the symmetry principles as well as the PLA are the GPN-laws. In contrast to Lange, I introduce the higher level of laws — the MN-laws that govern not only physical facts but also all kind of actual events.

So basically, it seems the PLA plays a unique role as one of the GPN-laws. It is an intermediate law between the LPN-laws and the MN-laws. Some laws of motion are direct mathematical consequences of the PLA, while others are not, but all are necessary consequences of the two MN-laws ruling the possible histories in the possible realm of our world.

To summarise the issue of the necessity of the PLA, I assert this principle as being a conceptual tool that enables us to deal theoretically with the actual world, although it lacks independent metaphysical essence (Humean view). At the same time, this principle is non-accidental, since the source of its necessity lies in the possible realm of the world. The PLA is the necessary consequence of two MN-laws (non-Humean view). Thus, the PLA lies between the Humean and non-Humean concepts.

## 10 Conclusion

In this paper, I have investigated the possible paths or possible histories in the PLA. I have examined connections of the possible histories with the metaphysical notion of "possibilia" and Leibniz's concept of the essences or possibles striving towards existence. Unfortunately, this concept has long been rejected by most philosophers; today, however, it finds an ally in quantum behaviour. Indeed, another source that inspires me is a deep relationship between the PLA and the quantum sum-over-histories model or the FPI.

In Section 6, I have elaborated the modal interpretation of the PLA based on two metaphysical models: modality and combination. According to the modality model, the possible histories in the PLA have essences in the possible modality but do not actually exist. Only actualised histories have existence in the actual modality. In compliance with the Leibnizian doctrine of the striving possibles, the essence of every possible history tends towards existence in our actual world. Using an analogy with the FPI, the modality model claims that all of the physical system's possible histories are jointly in the possible realm of our world. Therefore, this approach replaces the classical representation of a system's motion along a single actual history by a representation of simultaneous motions along an infinite set of all possible histories.

I have argued that, in the possible modality, a certain kind of a collision or "competition" occurs between all possible histories. The result of such a collision has to have the maximal essence and be manifested in the actual existence as the unique history. Other possible histories do not disappear completely as they still have essences, but they are not "compossible" in actuality. According to the combination model, the only actual history between any two actual events is the one that has the highest degree of the essence and combines the greatest number of possible histories between these events at the same time. From a physical view, it appears in the rule of the summation of the possible history's actions. Consequently, a metaphysical interpretation of the action is one of the physical measures of the essence, which consists of the necessity of each possible history to be realised in actuality.

The modal interpretation of the PLA changes the view of causality in this principle (Section 8). Indeed, a system does not need to "calculate" the value of the action or anything else. The rule of the summation of the history's actions does so, since the action is merely the physical measures of every history's essence. The system merely uses or actualises the maximal number of possibilities to move in each subsequent actual event. The continuous mutual play of a system's attempts in the possible modality and the following combining of these attempts create a system's actual events and actual history. Consequently, the significance of the PLA lies not with the mystic economy of nature, but merely in the observable effect of the collision and combination of all possible movements.

Such a modal approach to the PLA does not need the backward causation because, in the possible modality, the exact direction of time does not matter. It is also not contrary to the principle of locality and the relativistic limit of the speed of causal influences. In the possible realm of our world, there are not any physical restrictions. It means that all speeds of light and all time-like intervals are possible; therefore, all possible histories can occur simultaneously in all possible space-times.

To provide a new view of the PLA within the laws of nature, in Section 9, I introduced the system of three levels of laws. The LPN-laws describe only actual motions and follow the GPN-laws, which also involve the possible motions in the possible realm of our world and, in turn, are based on certain MN-laws. I consider the PLA is one of the GPN-laws. The PLA lacks independent metaphysical essence (Humean view); nevertheless it is the necessary consequence of the two MN-laws ruling the possible histories in the possible realm of our world (non-Humean view). Thus, the PLA lies between the Humean and non-Humean concepts.

The modal approach to the PLA can also explain the efficiency of the calculus of variations for a description of any kind of motion. It could explain why the various extremal principles are so widely distributed, not only in linear physics, but also in nonlinear thermodynamics, and biology. Perhaps, the universe does not "consider" its own states and histories successively, but all at once. Then the simple and constant rules of the combination of possibilities constitute the reason why our universe seems to us so uniform, ordered, and harmonious.

This conclusion entails a new view of the relationship between the PLA and quantum physics. In certain cases, the PLA can be considered as the classical limit of the FPI. However, this is just a mathematical derivation, as the FPI is unlikely to have an independent metaphysical essence (Humean view). Rather the deep ontological connection between the PLA and the FPI is made up in their possible histories, which obey the two MN-laws mentioned above (non-Humean view). There is one significant argument in favour of this hypothesis — a metaphysical interpretation of the action in the PLA (Section 6). I postulate that both the classical action and quantum action in the phase of probability amplitude to be merely different physical measures of the essence, which consist of the necessity of each possible history to be realised in actuality. The only difference is that the former relates to a classical phase or configuration space and the latter relates to a Hilbert space. Finally, both the variations in the PLA and the virtual path in the FPI can take place not only in a mathematician's head but also in the possible modality of being.

Summing up, I believe that amongst the various ways to explain the natural laws, the modal approach is the most promising one. It will definitely allow us to overcome a huge number of contradictions among the different fields of science, to understand how our world works and probably other worlds too.

## Acknowledgments

I am grateful to Hans Halvorson for his support and encouragement. He kindly read my paper and offered invaluable detailed comments and suggestions on the theme and organization of this paper. I would like to thank Michael Rota and Kirill Karpov for the chance to present a draft of this paper at the 2014 Paper Development Workshop in Analytical Philosophy and Theology–Russia. I would like to thank also two anonymous referees for their helpful feedback and criticism.